\documentclass[twocolumn,showpacs,preprintnumbers,amsmath,amssymb]{revtex4-1}

\usepackage{amsmath}
\usepackage{graphicx}
\usepackage{bm}% bold math

\begin{document}

\title{The exact relations between  the entanglement entropies of $XY$ and quantum Ising chains with free and fixed boundary conditions}

\author{Tianhao He} 

\author{Xintian Wu}

\affiliation{School of Physics and Astronomy, Beijing Normal University,
Beijing, 100875, China}

\email{Corresponding author: wuxt@bnu.edu.cn}

\date{\today}% It is always \today, today,

\begin{abstract}
The entanglement entropies of $XY$ chains and quantum Ising chains (QICs) with fixed boundary conditions are studied here. Three kinds of boundary conditions (BCs) are considered: fixed up--up or down--down (the spins at both ends are aligned in the same direction), fixed up--down or down--up (the spins at the two ends are aligned in opposite directions), and fixed--free (the spin at one end is aligned, and the other end is free). It is shown that i) the entanglement entropy of an $XY$ chain with a fixed--free BC is the sum of those of QICs with a fixed--free BC and with a free--free BC; ii) the entanglement entropy of an $XY$ chain with a fixed up--up 
boundary condition is the sum of those of QICs with a fixed up--up BC and with a free--free BC; and iii) the entanglement entropy of an $XY$ chain with a fixed up--down BC is the sum of that of a QIC with a fixed up--up BC and that of the first excited state of a QIC with a free--free BC. 
\end{abstract}

\maketitle

\section{INTRODUCTION}
Although sometimes viewed as an exotic phenomenon, entanglement is in fact generic and ubiquitous in quantum systems.  It is fundamental to quantum mechanics, and one of the features that distinguishes it sharply from classical mechanics. Given a way to divide a quantum system into two parts, and a state, it is natural to ask, “How entangled are they?” This turns out to be a very fruitful question in almost any area of physics where quantum mechanics plays a role, including the study of quantum many-body systems, quantum field theories, and even quantum gravity theories. In recent decades, entanglement has gone from being viewed as a quantum oddity to being seen as a quantifiable resource that be exploited to facilitate applications. Quantitative measures are particularly important in the context of quantum communication: some codes rely on sharing entangled quantum states and secrecy can only be guaranteed if the entanglement is sufficiently large.  The entropy of each subsystem quantifies how entangled they are (in a pure state), giving rise to the so-called entanglement entropy (EE). Entanglement thereby connects several areas of theoretical physics, including quantum information theory, condensed-matter theory, high-energy theory, and statistical mechanics \cite{witten,headrick,casini,calabrese}.

On one hand,  EE has been studied theoretically in recent two decades \cite{kitaev,latore}. In light of conformal field theory (CFT) \cite{cala,najafi,alba,cardy}, EE has been extensively studied for one-dimensional quantum spin models \cite{affleck,zhou,legeza,szirmai,taddia,fagotti,hu,tang}.

On the other hand , theoretical physicists have made great efforts to design the approaches to measure the entanglement entropy. Cardy proposed a method to measure the entanglement entropies in one-dimensional systems close to a quantum critical point in terms of the population of low-lying energy levels following a certain type of local quantum quench \cite{cardy}. Abanin et. al proposed a general method to measure the entanglement entropy \cite{abanin}. The method is based on a quantum switch (a two-level system) coupled to a composite system consisting of several copies of the original many-body system. The state of the switch controls how different parts of the composite system connect to each other. They showed that, by studying the dynamics of the quantum switch only, the Renyi entanglement entropy of the many-body system can be extracted. Santos et. al presented  a method to measure the von Neumann entanglement entropy of ground states of quantum many-body systems which does not require access to the system wave function \cite{dalmonte}. Sankar et. al propose a scheme to measure the topological entanglement entropy of  chiral topological phases, carrying protected edge states, based on a nontrivial connection with the  thermodynamic entropy change occurring in a quantum point contact (QPC) as it pinches off the topological liquid into two. We show how this entropy change can be extracted using Maxwell relations from charge detection of a nearby quantum dot \cite{sankar}.

Using new exact microscopic  numerical simulations of superfluid $^4He$, Herman et. al demonstrate for  the first time an area law scaling of entanglement entropy  ina real quantum liquid in three dimensions \cite{herman}. Tajik et. al measured the von Neumann entropy of spatially extended subsystems in an ultracold atom simulator of one-dimensional quantum field theories.  They experimentally verified one of the fundamental properties of equilibrium states of gapped quantum many-body systems—the area law of quantum mutual information. \cite{tajik} Chen et al.  observed the classical analogy of topological entanglement entropy experimentally \cite{chen}. The transverse Ising model can be simulated with cold atoms \cite{peter,ren}.

A major outcome of all these studies is 
%Editor: Please ensure that the intended meaning has been maintained in the following edit.
the ability to distinguish different phases and classify 
the critical points of continuous phase transitions into different universality classes according to EE. Most of the above studies were based on bulk properties; however, many studies have focused on EE in systems with boundaries. In the presence of boundaries, analytical and numerical calculations of EE are slightly more challenging because of the lack of translational invariance. Nevertheless, the EEs of a few quantum chains have been studied in the presence of boundaries with analytical and numerical techniques \cite{affleck,legeza,szirmai,castro,taddia,fagotti,affleck-1,xavier,wilzeck,kropin}.

Regarding the EEs of $XY$ chains and quantum Ising chains (QICs), it has been shown that there is an exact relationship between these EEs with periodic and open (free at both ends) boundary conditions (BCs) \cite{igloi}. In the same study, Igloi and Juhasz showed that an $XY$ chain with periodic and open BCs can be mapped to two QICs. Consequently, the EE of the $XY$ chain is the sum of the EEs of the two corresponding QICs.

In this work, we study three other types of conformal invariant boundary conditions (CIBCs). These CIBCs are 
%Editor: Please ensure that the intended meaning has been maintained in the following edit.
fixed--free (one end is fixed, and one end is free) and fixed--fixed BCs, where the fixed--fixed type can be $(++)$ (the spins at both ends are fixed in the same $x$-direction) or $(+-)$ 
(the spins at the two ends are fixed in opposite $x$-directions). Extending the reasoning of Igloi et al. \cite{igloi}, we find that i) the EE of an $XY$ chain with a fixed--free BC is the sum of those of QICs with a fixed--free BC and with a free--free BC; ii) the EE of an $XY$ chain with a fixed up--up boundary condition is the sum of those of QICs with a fixed up--up BC and with a free--free BC; and iii) the EE of an $XY$ chain with a fixed up--down BC is the sum of those of a QIC with a fixed up--up BC and the first excited state of a QIC with a free--free BC. 

This paper is arranged as follows. In Section II, the fixed BCs for $XY$ chains and QICs are discussed. In Section III, the relationships between the EEs of $XY$ chains and QICs are obtained. Section IV provides a summary. 

\section{Equivalent Hamiltonians for $XY$ and quantum Ising chains with fixed BCs}

For convenience, we consider a spin chain with $2L$ spins. The $XY$ model with boundary fields is defined as
\begin{equation}
H^{XY}_b=-\sum_{i=1}^{2L-1}  (J_i^xS_i^x S_{i+1}^x+J_i^yS_i^yS_{i+1}^y)-b_1S_1^x-b_{2L}S_{2L}^x
\label{eq:hamiltonx}
\end{equation}
Here, the $S_i^{x,y}$s are spin-1/2 operators at site i, and the couplings $J^x_i$ and $J_y^i$ may be different
and are site dependent in general. $b_1$ and $b_{2L}$ are the boundary fields. For $b_1=b_{2L}=0$, the boundary conditions are free. 

For $b_1=b_{2L}=\infty$, the spins at the two ends are aligned in the positive $x$ direction. Both ends have fixed boundary conditions. We denote 
%Editor: Please ensure that the intended meaning has been maintained in the following edit.
these BCs
 by``$(++)$". For $b_1=-b_{2L}=\infty$, the spins at the two ends are aligned in the positive and negative $x$ directions. We denote 
%Editor: Please ensure that the intended meaning has been maintained in the following edit.
this condition
 by``$(+-)$". If $b_1=\infty, b_{2L}=0$ (or $b_1=0,  b_{2L}=\infty$), we denote this by ``$(+f)$".

For the $(++)$ boundary condition, the eigen wavefunction $| \Psi \rangle$ satisfies
\begin{equation}
S_1^x | \Psi \rangle=S_{2L}^x | \Psi \rangle= \frac{1}{2} | \Psi \rangle
\end{equation}
This can be written as
\begin{equation}
|\Psi\rangle=|+\rangle_1 \otimes |\Psi'\rangle \otimes |+\rangle_{2L}
\label{eq:otimes++}
\end{equation}
where $S^x_1 |+\rangle_1 =\frac{1}{2}|+\rangle_1  $, $S^x_{2L} |+\rangle_{2L} =\frac{1}{2}|+\rangle_{2L}$, and $|\Psi'\rangle$ is the eigen wavefunction of the following Hamiltonian:
\begin{eqnarray}
H'^{XY}_{++} & = & -\sum_{i=2}^{2L-2}  (J_i^xS_i^x S_{i+1}^x+J_i^yS_i^yS_{i+1}^y) \nonumber \\ 
         &  & -\frac{J_1^x}{2}S_2^x-\frac{J_{2L-1}^x}{2}S_{2L-1}^x
\label{eq:++boundary}
\end{eqnarray}
The 
%Editor: Please ensure that the intended meaning has been maintained in the following edit.
two 
spins at the ends are effectively replaced by two boundary fields.

This wavefunction can be obtained in another way. Consider the following Hamiltonian:
\begin{eqnarray}
H^{XY}_{eff} & = & -\sum_{i=2}^{2L-2}  (J_i^xS_i^x S_{i+1}^x+J_i^yS_i^yS_{i+1}^y) \nonumber \\
                  &  & -J_1^xS_1^xS_2^x-J_{2L-1}^xS_{2L-1}^xS_{2L}^x.
\label{eq:hxyeff}
\end{eqnarray}
Note that the terms $S_1^y S_2^y$ and $S_{2L-1}^yS_{2L}^y$ are absent. Since the Hamiltonian $H^{XY}_{eff}$ commutes with $S_1^x$ and $S_{2L}^x$, it can be diagonalized in four sectors, labeled by the eigenvalues of $S_1^x$ and $S_{2L}^x$. We denote these sectors as ($\pm 1, \pm 1$), where $\pm$ are the signs of $2s_ j = \pm 1 $ ($j = 1$ or $j =2L$) and the eigenvalues of $S_1^x$ and $S_{2L}^x$.

In sector $(1,1)$, the eigen wavefunction is given by Eq. (\ref{eq:otimes++}), where $\Psi'$ also satisfies Eq. (\ref{eq:++boundary}). Setting $S_1^x= S^x_{2L}=\frac{1}{2}$ in Eq. (\ref{eq:hxyeff}) yields Eq. (\ref{eq:++boundary}). That is, one can obtain the eigen wavefunction for the fixed boundary condition $(++)$ of the Hamiltonian in Eq. (\ref{eq:hamiltonx}) with $b_1=b_{2L}=\infty$ by diagonalizing the Hamiltonian in Eq. (\ref{eq:hxyeff}) in sector $(1,1)$.

For the $(+-)$ boundary condition, the eigen wavefunction $| \Psi \rangle$ satisfies
\begin{equation}
S_1^x | \Psi \rangle=-S_{2L}^x | \Psi \rangle= \frac{1}{2} | \Psi \rangle
\end{equation}
This can be written as
\begin{equation}
|\Psi\rangle=|+\rangle_1 \otimes |\Psi'\rangle \otimes |-\rangle_{2L}
\label{eq:otimes+-}
\end{equation}
where $S^x_1 |+\rangle_1 =\frac{1}{2}|+\rangle_1  $, $S^x_{2L} |+\rangle_{2L} =-\frac{1}{2}|+\rangle_{2L}$, and $|\Psi'\rangle$ is the eigen wavefunction of the following Hamiltonian:
\begin{eqnarray}
H'^{XY}_{+-} & = & -\sum_{i=2}^{2L-2}  (J_i^xS_i^x S_{i+1}^x+J_i^yS_i^yS_{i+1}^y) \nonumber \\ 
         &  & -\frac{J_1^x}{2}S_2^x+\frac{J_{2L-1}^y}{2}S_{2L-1}^x
\label{eq:+-boundary}
\end{eqnarray}
The 
%Editor: Please ensure that the intended meaning has been maintained in the following edit.
two 
spins at the ends are effectively replaced by two boundary fields.

This wavefunction can be obtained in sector $(1,-1)$ of the Hamiltonian in Eq. (\ref{eq:hxyeff}). Setting $S_1^x=-S^x_{2L}=\frac{1}{2}$ in Eq. (\ref{eq:hxyeff}) yields Eq. (\ref{eq:+-boundary}). That is, one can obtain the eigen wavefunction for the fixed boundary condition $(+-)$ of the Hamiltonian in Eq. (\ref{eq:hamiltonx}) with $b_1=-b_{2L}=\infty$ by diagonalizing the Hamiltonian in Eq. (\ref{eq:hxyeff}) in sector $(1,-1)$.

For the $(+f)$ boundary condition, the eigen wavefunction $| \Psi \rangle$ satisfies
\begin{equation}
S_1^x | \Psi \rangle= \frac{1}{2} | \Psi \rangle
\end{equation}
This can be written as
\begin{equation}
|\Psi\rangle=|+\rangle_1 \otimes |\Psi'\rangle
\label{eq:otimes}
\end{equation}
where $S^x_1 |+\rangle_1 =\frac{1}{2}|+\rangle_1  $ and  $|\Psi'\rangle$ is the eigen wavefunction of the following Hamiltonian:
\begin{equation}
H'^{XY}_{+f}  = -\sum_{i=2}^{2L-1}  (J_i^xS_i^x S_{i+1}^x+J_i^yS_i^yS_{i+1}^y)-\frac{1}{2}J_1^xS_2^x
\label{eq:+fboundary}
\end{equation}
The 
%Editor: Please ensure that the intended meaning has been maintained in the following edit.
first 
spin at the left end is effectively replaced by the boundary field. Consider the following Hamiltonian:
\begin{equation}
H'^{XY}_{eff} = -\sum_{i=2}^{2L-1}  (J_i^xS_i^x S_{i+1}^x+J_i^yS_i^yS_{i+1}^y)-J_1^xS_1^xS_2^x.
\label{eq:eff+f}
\end{equation}
Setting $S_1^x=\frac{1}{2}$ in the above Hamiltonian yields the Hamiltonian in Eq. (\ref{eq:+fboundary}). Since the above Hamiltonian commutes with $S_1^x$, it can be diagonalized in two sectors, labeled by the eigenvalues of $S_1^x$. In terms of entanglement entropy, the two sectors are equivalent.

QICs with fixed boundary conditions can be dealt with similarly. Consider a QIC with length $L$:
\begin{equation}
H^I=-\frac{1}{2}{\big [ } \sum_{i=1}^{L-1}J_i \sigma^x_i \sigma^x_{i+1}+\sum_{i=1}^{L} \lambda_i  \sigma^z_i+b_1\sigma_1^x+b_L\sigma_L^x {\big ]}
\end{equation}
For $b_1=b_{L}=\infty$, the spins at the two ends are aligned in the positive $x$ direction. Both ends have a fixed boundary condition $(++)$. For $b_1=-b_{L}=\infty$, the spins at the two ends are aligned in the positive and negative $x$ directions. This is the fixed boundary condition $(+-)$. If $b_1=\infty, b_{L}=0$ (or $b_1=0,  b_{L}=\infty$), it is the fixed boundary condition $(+f)$.

For the fixed boundary conditions $(++)$ and $(+-)$, we can obtain the eigen wavefunction with the following Hamiltonian:
\begin{equation}
H^I_{eff}=-\frac{1}{2}{\big [ } \sum_{i=1}^{L-1}J_i \sigma^x_i \sigma^x_{i+1}+\sum_{i=2}^{L-1} \lambda_i  \sigma^z_i {\big ]}
\end{equation}
Since this commutes with $\sigma_1^x$ and $\sigma_{L}^x$, it can be diagonalized in four sectors, labeled by the eigenvalues of $\sigma_1^x$ and $\sigma_{L}^x$. We denote these sectors as ($\pm 1, \pm 1$), where $\pm$ are the signs of $s_ j = \pm 1 $ ($j = 1$ or $j =L$) and the eigenvalues of $\sigma_1^x$ and $\sigma_{L}^x$. As discussed for the $XY$ model, the eigen wavefunction with the fixed boundary condition $(++)$ can be obtained in sector $(1,1)$, and that with $(+-)$ can be obtained
in sector $(1,-1)$.

For the fixed boundary condition $(+f)$, consider the following Hamiltonian:
\begin{equation}
H'^I_{eff}=-\frac{1}{2}{\big [ } \sum_{i=1}^{L-1}J_i \sigma^x_i \sigma^x_{i+1}+\sum_{i=2}^{L} \lambda_i  \sigma^z_i {\big ]}.
\end{equation}
Since the term $\sigma_1^z$ is absent, this Hamiltonian commutes with $\sigma_1^x$; it has two sectors, labeled by the eigenvalues of $\sigma_1^x$. The eigen wavefunction with the fixed boundary condition $(+f)$ can be obtained in any sector.

In summary, the eigen 
%Editor: Please ensure that the intended meaning has been maintained in the following edit.
wavefunctions 
with fixed boundary conditions $(++)$ and $(+-)$ can be obtained in sectors $(1,1)$ and $(1,-1)$ for the Hamiltonian in Eq. (\ref{eq:hxyeff}); 
%Editor: Please ensure that the intended meaning has been maintained in the following edit.
that with the boundary condition $(+f)$ can be obtained in the sector in which the eigenvalue of $S^x_1$ is 
$\frac{1}{2}$ for the Hamiltonian in Eq. (\ref{eq:eff+f}). 

\section{Mapping from $XY$ chains to QICs}
Consider an $XY$ chain with length $2L$ with open boundaries:
\begin{equation}
H^{XY} =- \sum_{i=1}^{2L-1}  (J_i^xS_i^x S_{i+1}^x+J_i^yS_i^yS_{i+1}^y)
\label{eq:hxy}
\end{equation}

The three Hamiltonians can be diagonalized via the transformation \cite{lieb,pfeuty,vicari,bariev,hinrichsen,bilstein}
\begin{equation}
c_k=\sum_i ( g_{i,k}a_i+ h_{i,k} a^{\dagger}_i)
\end{equation}
with
\begin{equation}
g_{i,k}=\frac{\phi_{i,k}+\psi_{i,k}}{2}, ~~
h_{i,k}=\frac{\phi_{i,k}-\psi_{i,k}}{2}.
\end{equation}
The eigenvectors $\phi_{i,k}, \psi_{i,k}$ and the eigenvalues satisfy
\begin{equation}
[c_k, H]=\varepsilon_k c_k
\end{equation}

According to reference \cite{igloi}, the $XY$ chain with length $2L$ can be mapped into two isolated quantum Ising chains (QICs) with length $L$:
\begin{eqnarray}
H^{XY}  = \frac{1}{2} {\big [} b_{I(\sigma)}+b_{I(\sigma)} {\big ]} 
\end{eqnarray}
where
\begin{equation}
b_{I(\sigma)}=-\frac{1}{2}\sum_{i=1}^{L-1}J_{2i}^x \sigma^x_i \sigma^x_{i+1}-\frac{1}{2}\sum_{i=1}^{L} 
J_{2i-1}^y  \sigma^z_i
\label{eq:hsigma}
\end{equation}
and
\begin{equation}
b_{I(\tau)}=-\frac{1}{2}\sum_{i=1}^{L-1}J_{2i}^y \tau^x_i \tau^x_{i+1}-\frac{1}{2}\sum_{i=1}^{L}J_{2i-1}^x \tau_i^z
\label{eq:htau}
\end{equation}
They can also be diagonalized as in the $XY$ model \cite{lieb,pfeuty}. We denote the corresponding eigenvectors and eigenvalues by $\phi_{i,k}^{(\sigma)},\psi_{i,k}^{(\sigma)},\varepsilon_k^{(\sigma)}$ for the Hamiltonian $b_{I(\sigma)}$ and by $\phi_{i,k}^{(\tau)},\psi_{i,k}^{(\tau)},\varepsilon_k^{(\tau)}$ for the Hamiltonian $b_{I(\tau)}$.

Igloi and Juhasz found another way to solve this eigenvalue problem. For the $XY$ chain with length $2L$, the eigenvectors are $\phi_{i,k}, \psi_{i,k}$ with $i=1,2,\cdots,2L$. They constructed a new vector ${\bf u}=(\phi_{1,k},\psi_{1,k}$, where $\psi_{2,k},\phi_{2,k},\psi_{3,k},\phi_{3,k},\cdots , \phi_{2L-1,k},\psi_{2L-1,k},\psi_{2L,k},\phi_{2L,k})$. This vector is the eigenvector of the T-matrix of the following form:
\begin{eqnarray}
&T&_{XY} = \nonumber \\
& &\left ( \begin{array}{cccccccccc}
            0      &   0    &  J_1^y &          &     &    &    &    &  & \\
            0     &    0    & 0         &  J_1^x &     &    &   &    & & \\
    J_1^y      &   0    & 0         & 0         &  J_2^x  &    &   &   & &   \\
                  &    J_1^x    & 0 & 0 & 0 &  J_2^y &   &   &  &  \\
                 &      &    J_2^x    & 0 & 0 & 0 &  J_3^y &   & & \\
                 &      &    & J_2^y    & 0 & 0 & 0 & \ddots  &   & \\
                &      &    &     & \ddots & \ddots & \ddots & &J^y_{2L-1} &   \\
              &  &      &    &     & J_{2L-2}^y & 0 & 0 & 0  &J^x_{2L-1}    \\
               & &      &    &     &  & J_{2L-1}^y & 0 & 0  &0   \\
              &  &      &    &     &  &    & J_{2L-1}^x & 0 & 0    
           \end{array} \right )
\nonumber\\
\end{eqnarray}
That is,
\begin{equation}
\sum_{j=1}^{2L} (T_{XY})_{i,j}u_{j,k}=\varepsilon_k u_{i,k}
\end{equation}

The $2L$ eigenvectors of $T_{XY}$ with positive eigenvalues can be divided into two classes: 

i) The first class of vectors, which are marked with odd superscripts, 
%Editor: Please ensure that the intended meaning has been maintained in the following edit.
satisfy 
$\phi_{2i,2k-1}=\psi_{2i-1,2k-1}=0$, whereas the nonzero components of the vectors form vectors $(\phi_{1,2k-1},\psi_{2,2k-1} , \phi_{3,2k-1},\cdots,\phi_{2L-1,2k-1},\psi_{2L,2k-1})$. These are the eigenvectors of the following matrix:
\begin{equation}
T_{I}^{(\sigma)}=\left ( \begin{array}{cccccc}
            0    &  J_1^y   &           &             &    &       \\
    J_1^y     &    0        & J_2^x &             &    &   \\
                  &   J_2^x  & 0         & J_3^y  &    &    \\
                  &              &\ddots  & \ddots  & \ddots &    \\
                 &                &          & J_{2L-1}^x & 0 &J_{2L-1}^y   \\
                 &                &         &                     &J_{2L-1}^y  & 0     
           \end{array} \right ).
\end{equation}
This is simply the T-matrix of the Hamiltonian in Eq. (\ref{eq:hsigma}) \cite{igloi}. On the other hand, the eigenvector of the matrix $T_{I}^{(\sigma)}$ can be written as ${\bf u}^{(\sigma)}=(-\phi^{(\sigma)}_{1,k},\psi^{(\sigma)}_{1,k},-\psi^{(\sigma)}_{2,k},\cdots,-\phi^{(\sigma)}_{L-1,k},\psi^{(\sigma)}_{L,k})$, and it satisfies
\begin{equation}
\sum_{j=1}^L (T_{I}^{(\sigma)})_{i,j} u^{(\sigma)}_{j,k}=\varepsilon^{(\sigma)}_k u^{(\sigma)}_{i,k}
\end{equation}
Therefore, the following correspondence holds:
\begin{equation}
\varepsilon_{2k-1}=\varepsilon_k^{(\sigma)}
\end{equation}
and
\begin{equation}
\phi_{2i-1,2k-1}=-\phi^{(\sigma)}_{i,2k-1}, \hskip 0.3cm  \psi_{2i,2k-1}=\psi^{(\sigma)}_{i,2k-1}.
\label{eq:corr1}
\end{equation}

ii) The second class of vectors, which are marked with even superscripts, 
%Editor: Please ensure that the intended meaning has been maintained in the following edit.
satisfy
 $\phi_{2i-1,2k}=\phi_{2i,2k}=0$, and the nonzero components of the vectors form vectors $(\psi_{2,2k},-\phi_{1,2k},$ $\psi_{4,2k},\cdots,\psi_{2L,2k},-\phi_{2L-1,2k})$. These are the eigenvectors of the following matrix:
\begin{equation}
T_{I}^{(\tau)}=\left ( \begin{array}{cccccc}
           0    &  J_1^x   &           &             &    &       \\
    J_1^x     &    0        & J_2^y &             &    &   \\
                  &   J_2^y  & 0         & J_3^x  &    &    \\
                  &              &\ddots  & \ddots  & \ddots &    \\
                 &                &          & J_{2L-1}^y & 0 &J_{2L-1}^x   \\
                 &                &         &                     &J_{2L-1}^x  & 0     
           \end{array} \right ).
\end{equation}
This is simply the T-matrix of the Hamiltonian in Eq. (\ref{eq:htau}) \cite{igloi}. Its eigenvector can be written as ${\bf u}^{(\tau)}=(-\phi^{(\tau)}_{1,k},\psi^{(\tau)}_{1,k} , -\phi^{(\tau)}_{2,k},\cdots,-\phi^{(\tau)}_{L-1,k},\psi^{(\tau)}_{L,k})$, and it satisfies
\begin{equation}
\sum_{j=1}^L (T_{I}^{(\tau)})_{i,j} u^{(\tau)}_{j,k}=\varepsilon^{(\tau)}_k u^{(\tau)}_{i,k}
\end{equation}
with the correspondence
%Editor: Please ensure that the intended meaning has been maintained in the following edit. Text was deleted because it presented nearly identical information.

\begin{equation}
\varepsilon_{2k}=\varepsilon_k^{(\tau)}
\end{equation}
\begin{equation}
\phi_{2i,2k}=\psi^{(\tau)}_{i,2k}, \hskip 0.3cm  \psi_{2i-1,2k}=-\phi^{(\tau)}_{i,2k}
\label{eq:corr2}
\end{equation}

For free-fermionic systems, the entanglement entropy of a subsystem with length $2l$ can be obtained from the restricted correlation matrix \cite{peschel,latore}, the elements of which are
given by
\begin{eqnarray}
G_{m,n} & = & \langle 0|(a^{\dagger}_n-a_n)(a^{\dagger}_m-a_m)|0\rangle \nonumber \\
               & = & -\sum_{k=1}^{2L}\psi_{m,k}\phi_{n,k}, \hskip 0.2cm m,n=1,2,\cdots,2l
\end{eqnarray}
Using the properties of $\phi_{i,k},\psi_{i,k}$ and the correspondence in Eqs. (\ref{eq:corr1}) and (\ref{eq:corr2}), one can obtain
\begin{eqnarray}
G_{2i-1,2j-1}& = & 0, \hskip 1cm G_{2i-1,2j} = -G^{(\sigma)}_{i,j} \nonumber \\
G_{2i,2j-1}& = & -G^{(\tau)}_{i,j}, \hskip 1cm G_{2i,2j}= 0
\end{eqnarray}
where $G^{(\sigma,\tau)}_{ i,j}$ denotes the matrix element of the correlation matrix of the QIC with the Hamiltonian $b_{I (\sigma,\tau)}$. The correlation matrix for 
%Editor: Please ensure that the intended meaning has been maintained in the following edit.
even superscripts 
is bipartite and is composed of $2 \times 2$ matrices
\begin{equation}
\left ( \begin{array}{cc}
         0    &   -G^{(\sigma)}_{i,j}         \\
         -G^{(\tau)}_{i,j}    & 0  
         \end{array} \right ).
\end{equation}

The entanglement entropy is given by
\begin{equation}
S^{(XY)}(2l,2L)=-\sum_{k=1}^{2l}{\Big [} \frac{1-\nu_k}{2} \ln \frac{1-\nu_k}{2}+\frac{1+\nu_k}{2} \ln \frac{1+\nu_k}{2} {\Big ]}
\end{equation}
where $\nu_k$ is the eigenvalue of the matrix ${\bf G} {\bf G} ^T$. For an XY chain with even $2l$,
 %the matrix ${\bf G} {\bf G} ^T$ is composed of 2×2 diagonal matrices
\begin{equation}
\left ( \begin{array}{cc}
         [{\bf G} ^{(\sigma)} {\bf G} ^{(\sigma)T}]_{i,j}    &  0           \\
         0    & [{\bf G} ^{(\tau)}{\bf G} ^{(\tau)T}]_{i,j}   
         \end{array} \right ).
\end{equation}
Thus, ${\bf G} {\bf G} ^T$ can be written in a form that consists of two diagonal blocks, and the eigenvalues are obtained by solving the two separate eigenvalue problems of ${\bf G} ^{(\sigma)} {\bf G} ^{(\sigma)T}$ and ${\bf G} ^{(\tau)}{\bf G} ^{(\tau)T}$.
From the above relations, Igloi and Juhasz proved that the entanglement entropy of the XY chain in Eq. (\ref{eq:hxy}) is the sum of the entanglement entropies of the two QICs defined in Eqs. (\ref{eq:hsigma}) and (\ref{eq:htau}) \cite{igloi}:
\begin{equation}
S^{(XY)}(2l,2L)=S^{(\sigma)}(l,L)+S^{(\tau)}(l,L).
\label{eq:xy-ising}
\end{equation}

\section{Exact relationship between $XX$ chains and QICs with fixed boundaries}
In the previous section, the discussion concerns general cases. In this section, we discuss homogeneous critical $XX$ (isotropic $XY$) chains with fixed--free, fixed up--up and 
%Editor: Please ensure that the intended meaning has been maintained in the following edit.
fixed 
up--down BCs. These models are extensively studied numerically and with CFT.

For fixed--free BCs, the homogeneous critical $XX$ Hamiltonian is given by Eq. (\ref{eq:+fboundary}) with $J_i^x=J_i^y=1, i=2,3,\cdots,2L$ and $J_1^x=1$. That is,
\begin{equation}
H'^{XX}_{eff}  =- \sum_{i=2}^{2L-1}  (S_i^x S_{i+1}^x+S_i^yS_{i+1}^y)- S_1^x S_2^x
\end{equation}
According to the mapping from Eq. (\ref{eq:hxy}) to Eq. (\ref{eq:hsigma}) and Eq. (\ref{eq:htau}), we obtain the corresponding Hamiltonians
\begin{equation}
H^{I(\sigma)}=-\frac{1}{2}\sum_{i=1}^{L-1}\sigma_i^x \sigma_{i+1}^x -\frac{1}{2}\sum_{i=2}^L \sigma_i^z,
\end{equation}
which has $(+F)$ BCs according to Eq. (\ref{eq:eff+f}), and
\begin{equation}
H^{I(\tau)}=-\frac{1}{2}\sum_{i=1}^{L-1}\tau_i^x \tau_{i+1}^x -\frac{1}{2}\sum_{i=1}^L \tau_i^z,
\end{equation}
which has free--free BCs. According to Eq. (\ref{eq:xy-ising}), we obtain
\begin{equation}
S_{+f}^{(XX)}(2l,2L)=S^{(\sigma)}_{+f}(l,L)+S^{(\tau)}_{ff}(l,L).
\label{eq:xy-ising+f}
\end{equation}

For the fixed up--up $(++)$ and up--down $(+-)$ BCs, we need to consider the Hamiltonian defined in Eq. (\ref{eq:hxyeff}) with $J_i^x=J_i^y=1, i=2,3,\cdots,2L-2$ and $J_1^x=J^x_{2L-1}=1$. That is,
\begin{equation}
H'^{XY}_{eff}  = -\sum_{i=2}^{2L-2}  (S_i^x S_{i+1}^x+S_i^yS_{i+1}^y)+ S_1^x S_2^x+ S_{2L-1}^x S_{2L}^x
\label{eq:hxypure}
\end{equation}
According to the mapping from Eq. (\ref{eq:hxy}) to Eq. (\ref{eq:hsigma}) and Eq. (\ref{eq:htau}), we obtain the corresponding Hamiltonians
\begin{equation}
H^{I(\sigma)}=-\sum_{i=1}^{L-1}\sigma_i^x \sigma_{i+1}^x -\sum_{i=2}^{L-1} \sigma_i^z,
\label{eq:hsigmapure}
\end{equation}
which has four sectors corresponding to the $(++)$, $(+-)$, $(-+)$ and $(--)$ BCs according to Eq. (\ref{eq:eff+f}), and
\begin{equation}
H^{I(\tau)}=-\sum_{i=1}^{L-1}\tau_i^x \tau_{i+1}^x -\sum_{i=1}^L \tau_i^z,
\label{eq:htaupure}
\end{equation}
which has free--free BCs. Generally, the entanglement entropy is referred to as the ground state. The ground states of $H'^{XY}_{eff}$ belong to sectors $(1,1)$ 
%Editor: Please ensure that the intended meaning has been maintained in the following edit.
and
 $(-1,-1)$. The two corresponding QICs should also be in the ground state. According to Eq. (\ref{eq:xy-ising}), we obtain
\begin{equation}
S^{(XX)}_{++}(2l,2L)=S^{(\sigma)}_{++}(l,L)+S^{(\tau)}_{ff}(l,L).
\label{eq:xy-ising++}
\end{equation}

The ground state of the $XX$ chain with a fixed up--down $(+-)$ boundary condition belongs to sector $(1,-1)$, the Hamiltonian defined in Eq. (\ref{eq:hxypure}), and it is its first excited state \cite{xavier}. Which state does this excited state correspond to in the two QICs? Is it the first excited state for $H^{I(\sigma)}$ defined in Eq. (\ref{eq:hsigmapure})? Or is it the first excited state for $H^{I(\tau)}$ defined in Eq. (\ref{eq:htaupure})? We will show that the energy for the first excited state of $H^{I(\tau)}$ defined in Eq. (\ref{eq:htaupure}) is lower than that of $H^{I(\sigma)}$ defined in Eq. (\ref{eq:hsigmapure}).

According to reference \cite{lieb,vicari}, the eigenvector $\psi^{(\sigma)}$ is given by
\begin{equation}
\psi^{(\sigma)}_k (n-1)+ 2\psi^{(\sigma)}_k (n ) +\psi^{(\sigma)}_k (n+1)  = (\varepsilon_k^{(\sigma)})^2  \psi^{(\sigma)}_k (n)
 \label{eq:tauvector} 
\end{equation}
for $ ~~ 1<n<L-1$, and at the two boundaries,
\begin{eqnarray}
\psi^{(\sigma)}& ( & 1) +\psi^{(\sigma)}(2 ) =(\varepsilon_k^{(\sigma)})^2  \psi^{(\sigma)}(1)  \nonumber  \\
\psi^{(\sigma)}& ( & L-2)  + \psi^{(\sigma)}(L-1)  =(\varepsilon_k^{(\sigma)})^2  \psi^{(\sigma)}(L-1)   \nonumber \\
\psi^{(\sigma)}& ( & L)  = 0 
\label{eq:sigmab-1}
\end{eqnarray}
The eigenvector is assumed to be given by
\begin{equation}
\psi^{(\sigma)}_k (n)=(-1)^n A\sin (nk+\delta)
\end{equation}
From Eq. (\ref{eq:tauvector}), we obtain
\begin{equation}
\varepsilon_k^{(\sigma)}=\sqrt{2(1-\cos k)}
\end{equation}
To satisfy the boundary conditions in Eq. (\ref{eq:sigmab-1}), we set
\begin{equation}
k=\frac{(2n-1)\pi}{L-3/2},~~~~n=1,2,3,\cdots
\end{equation}
The first excited state has the eigenenergy
\begin{equation}
\varepsilon^{(\sigma)}_1=\sqrt{2(1-\cos \frac{\pi}{L-3/2})}
\end{equation}

Similarly, $\psi^{(\tau)}$ is given by
\begin{equation}
\psi^{(\tau)}_k (n-1)+ 2\psi^{(\tau)}_k (n)  +\psi^{(\tau)}_k (n+1)  = (\varepsilon_k^{(\tau)})^2  \psi^{(\tau)}_k (n)
   \label{eq:tauvector} 
\end{equation}
 for $1<n<L$, and at the two boundaries,
\begin{eqnarray}
2\psi^{(\tau)}_k (1) &+ &\psi^{(\tau)}_k (2)  =(\varepsilon_k^{(\tau)})^2  \psi^{(\tau)}_k (1)  \nonumber  \\
\psi^{(\tau)}_k (L-1) & + &2\psi^{(\tau)}_k (L)  =(\varepsilon_k^{(\tau)})^2  \psi^{(\tau)}_k (L) \label{eq:taub-1}  
\end{eqnarray}
The eigenvector is assumed to be given by
\begin{equation}
\psi_k^{(\tau)}(n)=(-1)^n A\sin (nk+\delta)
\end{equation}
From Eq. (\ref{eq:tauvector}), we obtain
\begin{equation}
\varepsilon_k^{(\tau)}=\sqrt{2(1-\cos k)}
\end{equation}
To satisfy the boundary conditions in Eq. (\ref{eq:sigmab-1}), we find that
\begin{equation}
k=\frac{(2n-1)\pi}{L-1},~~~~n=1,2,3,\cdots
\end{equation}
The first excited state has the eigenenergy
\begin{equation}
\varepsilon^{(\tau)}_1=\sqrt{2(1-\cos \frac{\pi}{L-1})}
\end{equation}

Since $\varepsilon^{(\tau)}_1<\varepsilon^{(\sigma)}_1$, the first excited state of the Hamiltonian in Eq. (\ref{eq:hxypure}) is mapped to the first excited state of the Hamiltonian in Eq. (\ref{eq:htaupure}) and the ground state of the Hamiltonian in Eq. (\ref{eq:hsigma}). Therefore, we have
\begin{equation}
S^{(XX)}_{+-}(2l,2L)=S^{(\sigma)}_{++}(l,L)+S^{(\tau)exc}_{ff}(l,L),
\label{eq:xy-ising+-}
\end{equation}
where $S^{(\tau)exc}_{ff}(l,L)$ is the EE for the first excited state of the Hamiltonian in Eq. (\ref{eq:htaupure}).

%The entanglement $S^{(XY)}(2l,2L)$ is obtained from the reduced density matrix 
%$\rho_{2l}=Tr_{2L-2l} |0\rangle_{XY} \langle 0|_{XY}$
%of  a block of length $2l$ consisting of spins $i=1,2,\cdots,2l$.  The entanglement entropies $S^{(\sigma)}(l,L)$ and $S^{(\tau)}(l,L)$ are obtained from the reduced density matrixes of the  blocks of length $l$ with corresponding spins $i=1, 2, \cdots ,l$.

\section{summary}
In this work, we derived the exact  relations in Eqs. (\ref{eq:xy-ising+f}), (\ref{eq:xy-ising++}) and (\ref{eq:xy-ising+-}) between the entanglement entropy of the XY chain and that of the QIC with free and fixed BCs. These relations are valid for a finite block of even size and hold for inhomogeneous couplings. We have carried out numerical calcultion and prove these relations \cite{he}. Since the derivation is based on a mapping between two correlation matrices, similar relations can be obtained for other measures of entanglement, such as Renyi entropy or concurrence.

\end{document}